\documentclass[twocolumn,amsmath,amssymb,floatfix]{revtex4}

\usepackage{bm}
\usepackage{amsmath}
\usepackage{epsf}

\newcommand{\beq}{\begin{equation}}

\newcommand{\eeq}{\end{equation}}
\newcommand{\beqa}{\begin{eqnarray}}
\newcommand{\eeqa}{\end{eqnarray}}

\begin{document}
\bibliographystyle{apsrev}
\title{Crossing the Phantom Divide: Dark Energy Internal Degrees of Freedom}
\author{Wayne Hu}
\affiliation{
Kavli Institute for Cosmological Physics, Department of Astronomy \& Astrophysics,
Enrico Fermi Institute, University of Chicago, Chicago IL 60637}

\begin{abstract}
\baselineskip 11pt
Dark energy constraints have forced viable alternatives that differ substantially
from a cosmological constant $\Lambda$
to have an equation of
state $w$ that {\it evolves} across the phantom divide set by $\Lambda$. 
Naively, crossing this divide makes the dark energy gravitationally unstable, a problem
that is typically finessed by unphysically ignoring the perturbations. 
While this procedure does not affect constraints near the favored cosmological constant model it
can artificially enhance the confidence with which alternate models are rejected.  Similar
to the general problem of stability for $w< 0$, the solution lies in the internal degrees of
freedom in the dark energy sector.  We explicitly show how to construct a two scalar field model that
crosses the phantom divide and mimics the single field behavior on either side to substantially
better than 1\% in all observables.  It is representative of models
 where the internal degrees
of freedom  keep the dark energy smooth out to the horizon scale independently
of the equation of state. 
\end{abstract}
\maketitle

\section{Introduction}

A self-consistent model for the dark energy requires not only a parameterization of
the evolution of its equation of state in the background $w=p/\rho$ but also a physical model for
its spatial fluctuations to guarantee gravitational stability.
Cosmological constraints on a constant equation of state have continued to close in upon $w=-1$
(e.g. \cite{Seletal04}).  
Since models with $w<-1$ are called phantom dark energy models (e.g. \cite{Cal02}),
we call this the ``phantom divide".  Viable alternate models for a
strongly 
evolving $w$ therefore must cross the divide at intermediate redshift so that
the effects on either side cancel.
   Simple
generalizations
of a single scalar field model that cross the divide  in fact cause severe gravitational instabilities in the
dark energy sector (e.g. \cite{Vik04}).

The usual approach in the literature for dealing with such cases 
is to artificially turn off the dark energy perturbations.    Doing so violates energy-momentum
conservation whenever $w \ne -1$.   
The justification for dropping these
perturbations
is that observations already place $w$ close to $-1$ and so the induced error is presumably small in
some physical realization of a crossing model.  
While true for the currently allowed $1\sigma$ deviations of $\Delta w \sim 0.1$,  the confidence level
at which larger deviations can be rejected can be affected.   Furthermore, a strong time evolution
allows $w$ to differ substantially from $-1$ during some epochs and still be consistent
with the distance data. It is therefore important to show explicitly that models exist where the dark energy remains
smooth as it crosses the phantom divide as implied by the usual procedure.

The need for a self-consistent treatment of the evolution of the dark energy is most 
apparent for the cosmic microwave background.
Here the ISW effect is sensitive to the
decay of the gravitational potential and, for example, the difference between the smooth and
clustered regimes of the dark energy for a constant $w\approx -2/3$
is roughly $50\%$.     While the impact for a canonical
 scalar field which is smooth out to the horizon scale is lower, it is well known \cite{CalDavSte98}
 that it changes CMB predictions
 significantly for larger $w$ and hence the confidence level of constraints on highly deviant
 $w$ (e.g. \cite{WelLew03}).
 
 In this Brief Report, we explicitly construct a two scalar field model of the dark energy that is gravitationally
 stable across the phantom divide and matches the canonical single scalar field predictions on
 either side to much better than $1\%$ in all observables.  
 Taken literally, such a model of course compounds the coincidence problem of dark energy
 but as we shall show it is a proxy for a potentially wider class of dark energy 
 models whose internal degrees of freedom keep it smooth out to the horizon scale. 

\section{Instability}
\label{sec:instability}

It is well known that dark energy models beyond a cosmological constant  require
internal degrees of freedom, or the presence of  non-adiabatic stress perturbations,
to remain gravitationally stable.  This necessity arises from  momentum conservation.
 Consider the dimensionless momentum
density $\rho u_i = T^0_{\hphantom{0}i}$ of the dark energy stress tensor $T^{\mu}_{\phantom{\mu}\nu}$.  The scalar
component in Fourier
space (e.g. \cite{Bar80} and \cite{GorHu04} for a pedagogical treatment in the same
notation)
\begin{equation}
\rho \dot u = {\dot a \over a}(3w-1)\rho u + k\delta p + (1+w)\rho k A\,,
 \end{equation}
 where $2A=\delta g_{00}/ g_{00}$ is the time-time perturbation to the metric in an arbitrary
 gauge.
 Given dark energy fluctuations that are internally adiabatic $p(\rho)$ 
 (not to be confused with adiabatic across all
 energy density components)
\begin{align*}
\delta p &=  \left( {\dot p \over \dot \rho} \right) \delta \rho = \left( w - {1\over 3}{d\ln (1+w) \over d\ln a} \right)\delta\rho \,,
\label{eqn:adiabaticpressure}
\end{align*}
where $w = p/\rho$ in the background.
Perturbations go unstable whenever the pressure response to a density fluctuation
is negative or singular.  The former occurs for adiabatic pressure fluctuations if
 $w <0$ and $1+w$ is slowly
varying.  A viable dark energy candidate in this
regime must contain internal degrees of freedom that supply a non adiabatic
pressure.  For a single scalar field with a canonical or non-canonical
kinetic term, this is achieved through separate kinetic and potential contributions 
to the energy density and pressure.

 In this case, one defines a sound speed
$c_e^2$ which relates the energy density and pressure fluctuation of the
kinetic term \cite{GarMuk99} or equivalently of the zero momentum or constant field gauge 
\cite{Hu98} and  obtains
\begin{equation}
\delta p = c_e^2 \delta \rho + 3{\dot a \over a}{1\over k}(c_e^2 - {\dot p \over \dot \rho} ) \rho u\,.
\label{eqn:kessence}
\end{equation}
If the sound speed $c_e\le 1$ it determines the scale under which the
dark energy is effectively smooth through the sound horizon $\int c_{e} da/(a^{2}H)$. 
For the canonical kinetic term $c_e=1$.  More specifically, well above this scale, stress gradients
are negligible and the gravitational potential $A=\Psi$ in the Newtonian gauge evolves as
\begin{equation}
\Psi \propto \left( 1 - {H(a) \over a} \int_{0}^a {da ' \over H(a')} \right) \,.
\label{eqn:clustered}
\end{equation}
Well below this scale the dark energy is smooth and $\Psi \propto G$ with
\begin{eqnarray}
\frac{d^2 G}{d \ln a^2}  + \left[ \frac{5}{2} - \frac{3}{2} w(a) \Omega_{\rm DE}(a) \right]
\frac{d G}{d \ln a} && \nonumber\\
 + 
\frac{3}{2}[1-w(a)]\Omega_{\rm DE}(a) G &=&0\,,
\label{eqn:smooth}
\end{eqnarray}
where $\Omega_{\rm DE}(a) = 8\pi G\rho/3 H(a)^2$ and $a=1$ is assumed where no
dependence is given. 
Note that in both limits, the effect of the dark energy on the potential is solely a function of
its background energy density.  The true degree of freedom in a dark energy model is where this
transition occurs.  Any physical solution to the instability problem that matches a desired
$w(a)$ and transition scale will be a fairly robust representation of the class of models.
We will use this fact below to replace the usual single scalar field ansatz with two scalar
fields.
 
A single scalar field does not generally solve the problem that $\dot p/\dot \rho$ becomes
singular as $w$ evolves across the phantom divide 
$w=-1$ with finite slope $d (1+w)/d\ln a \ne 0$ since $\dot p/\dot\rho$ still appears in
Eqn.~(\ref{eqn:kessence}).  Stability can obviously be achieved by an alternate ansatz for the internal degrees of freedom.  The simplest solution that
preserves the behavior of the single scalar field transition scale away from the crossing
point is to introduce multiple scalar fields. 

\section{Two Field Model}
\label{sec:twofield}

For definiteness,
the target form of $w(a)$ that we wish to model with two scalar fields is
\begin{equation}
w(a) = w_0 + (1-a) w_a \,,
\label{eqn:true}
\end{equation}
where $w_{a}$ is a constant.   For simplicity, we will
take the two fields, denoted ``$+$" and ``$-$" to individually  have constant equations of state
\begin{align}
w_\pm & = \bar w \pm \epsilon_n\,. 
\end{align}
Note that this restriction to strictly constant equations of state is not essential to
the construction.
Though a strictly constant $w$ for a rolling scalar field is unrealistic, 
scalar field potentials do exist where the resultant
equation of state differs significantly from $\pm 1$ and roughly constant during
the redshifts of interest
\cite{SteWanZla99}.   In any case the point of this explicit construction is
to provide a crossing model that is simple to implement in existing cosmological
codes and does not violate energy-momentum conservation.  It is not intended
to be a well-motivated model.

The equations of state define the relative energy density contributions
as a function of redshift or scale factor
\begin{align}
{\rho_\pm \over \rho_{\rm DE}}(a)&= {1 \over 2} [1\pm \delta(a)]\,, 
\end{align}
where
\begin{align}
\delta(a) = \delta_n + { (1-\delta_n^2) [1-(a/a_n)^{6\epsilon_n}]
\over (1+\delta_n) + (1-\delta_n) (a/a_n)^{6\epsilon_n} }\,.
\end{align}
Here $\delta_n = \delta(a_n)$ defines the ratio at a normalization epoch $a_n$.  This epoch
 should {\it not} be chosen as $a_n=1$ since variations in the equation of state locally leave
no net effect.  Rather $a_n=3/4$ is roughly the pivot point where variations in the
equation of state make the maximal effect on the high and low redshift observables.
 We will adopt this value as the matching point between the two field
model and the target $w(a)$.

With equal sound speeds in the two components, the pressure fluctuation becomes
\begin{equation}
\delta p = c_e^2 \delta \rho + 3 {\dot a \over a}{1\over k}\left[ c_{e}^{2}{\rho u}
- w_{+} \rho_{+}u_{+} - w_{-}\rho_{-}u_{-}\right]\,,
\end{equation}
Here $\rho=\rho_{+}+\rho_{-}$, $p = p_{+}+p_{-}$, 
$\rho u  = \rho_{+}u_+ + \rho_{-}u_-$
none of which contain singularities at the crossing.

The model dark energy equation of state becomes
\begin{align}
w_{\rm mod}(a) = \bar w + \delta(a) \epsilon_n
\end{align}
and at the normalization point $w_n  = \bar w + \delta_n \epsilon_n$.
Likewise at the normalization point the derivative of $w$ is 
\begin{align}
w_a = - {d w_{\rm mod} \over da} \Big|_{a_n} 
={ 3 \over a_n} \epsilon_n^2 (1-\delta_n^2) \,.
\label{eqn:wa}
\end{align}
Thus $w_{a}>0$ if $\delta_{n}^{2}<1$.  A negative derivative is possible if
$\delta_{n}^{2}>1$, i.e. if the $\rho_{-}<0$.

The two field model has 3 parameters $\bar w$, $\epsilon_{n}$ and $\delta_{n}$.  The
target model has 2 parameters $w_{0}$ and $w_{a}$, leaving one adjustable parameter
to improve the performance of the parameterization.  Choosing this parameter to
be $\epsilon_{n}$ defines the other two in terms of $w_{0}$, $w_{a}$ and $\epsilon_{n}$
\begin{align}
\bar w = w_0 + (1-a_n) w_a - \delta_n \epsilon_n \,,
\end{align}
and $\delta_{n}$ 
\begin{align}
\delta_n  = \left[ 1 - { w_a a_n \over 3 \epsilon_n^2} \right]^{1/2}\,.
\end{align}

We now choose $\epsilon_{n}$ to satisfy several criteria.  
Firstly, as  $w_a \rightarrow 0$ the two field model should reduce to a single field so as to
exactly match the standard dynamics.  Secondly, the two field model should match a
sufficiently
wide range of $-w_{\rm max} < w_a<w_{\rm max}$. We will take $w_{\rm max}=1$.
  Finally, it should minimize higher order
derivatives.

To satisfy the first condition, $\epsilon_{n}$ should scale as a power of $w_{a}$ at least near
$w_{a}=0$. Combined with the second criteria,
\begin{align}
\epsilon_n = \left( { a_n w_{\rm max} \over 3 }\right)^{1/2} \Big|{ w_a \over w_{\rm max}} \Big|^{(1-p)/2}\,,\quad (w_{a}>0)
\end{align}
since $w_{a}$ is maximized at $\delta_{n}=0$.

The final condition is that higher order derivatives such as
\begin{align}
w_{aa} \equiv {d^2 w \over da^2} = {1 -6 \delta_n \epsilon_n \over a_n}w_a
\end{align}
should be minimized in so far as possible. At large $w_{a}$ this is ensured by keeping
$\delta_{n}\epsilon_{n}$ small.  A good choice to minimize these higher order derivatives
is to take $p=1/10$.
For $w_{a}<0$, $\delta_{n}>1$ and so one must
restrict the values of $\epsilon_{n}$.  We take
\begin{equation}
\epsilon_{n} \rightarrow {\rm max}(\epsilon_{n},0.1)\,, \quad (w_{a}<0)\,.
\end{equation}
Since $1-\delta_{n}^{2}$ is unbounded from below, this restriction still allows large
negative $w_{a}$ in Eqn.~(\ref{eqn:wa}) while preserving the single field correspondence for $w_{a} \rightarrow 0$
from below.

\begin{figure}[tb]
\centerline{\epsfxsize=3.0in\epsffile{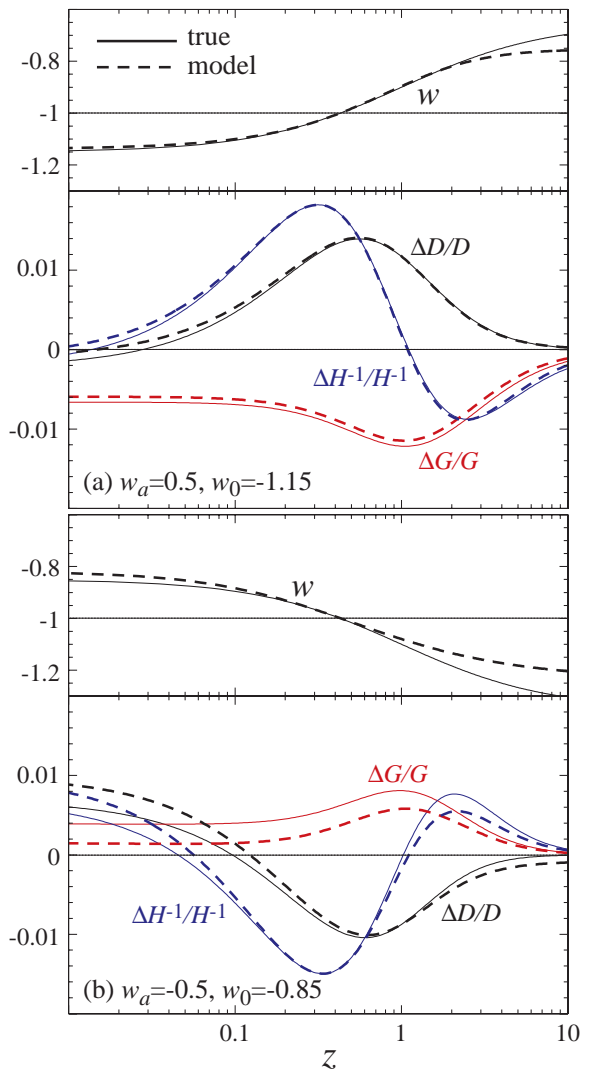}}
\caption{Target versus 2 field model behavior of $w$ and the dark energy observables relative
to a fiducial $\Lambda$ model (dotted line, see text).  Even in this nearly worst case scenario where $w$ crosses
the phantom divide near the pivot point $a \sim 3/4$ or $z\sim 0.3$, the model reproduces
the observables to $0.1\%$ accuracy for $w_a>0$ and only a factor of a few larger for $w_a<0$.}
\label{fig:time}
\end{figure}

To complete the modelling we match the {\it physical} energy density (i.e. not
relative to critical) at the normalization epoch $a_n=3/4$.
To scale the dark energy density of the target dark energy model Eqn.~(\ref{eqn:true})
we write $\rho_{\rm DE}(a_n)  = \Omega_{\rm DE} \rho_{\rm crit} g_{\rm DE}$ where
\begin{align}
g_{\rm DE} &= a_n^{-3(1+w_0+w_a)}e^{-3(1-a_n)w_a}
\end{align}
and analogously introduce
\begin{align}
g_\pm = a_n^{-3(1+w_\pm)}\,.
\end{align}
Thus given a target model with an energy density relative to critical of $\Omega_{\rm DE}$ and
Hubble constant parameterized by $H_0 = 100 h$km s$^{-1}$ Mpc$^{-1}$, the 2 field model
has an effective Hubble constant $h_{\rm e}$ 
\begin{align}
\left( { h_{\rm e} \over h }\right)^2 &=
\Omega_{\rm DE} {g_{\rm DE}} \left( { 1+\delta_n \over 2 g_+} + {1 -\delta_n \over 2g_-}\right) +1-\Omega_{\rm DE}
\end{align}
and density relative to critical defined by that Hubble constant of
\begin{align}
\Omega_{\pm} &= \Omega_{\rm DE} g_{\rm DE} {1\pm \delta_n\over 2g_\pm}\left( {h_{\rm e} \over h} \right)^{-2}\,.
\end{align}

Although this construction is completely phenomenological and hence physically contrived, the general point is that once a close
matching of $w(a)$ to some target 
has been achieved with multiple scalar fields, the scalar field dynamics
will make the predictions in both the smooth and clustered regimes robust to reparameterization.

\begin{figure}[tb]
\centerline{\epsfxsize=3.0in\epsffile{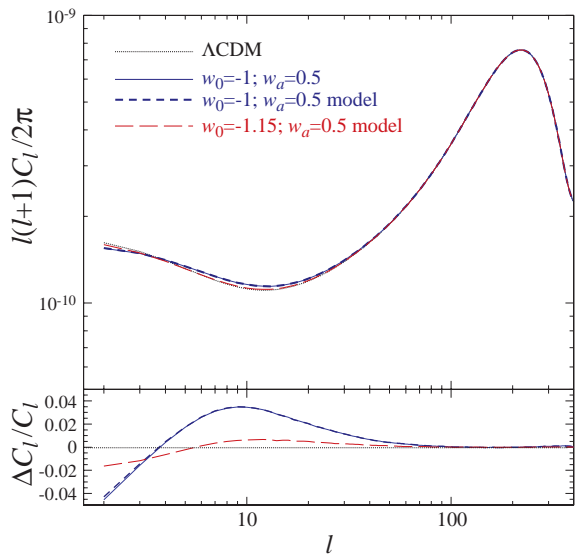}}
\caption{Model predictions for the CMB anisotropy power 
spectrum relative to a fiducial
$\Lambda$CDM model.   For models that do not cross the
phantom divide, e.g. $w_0=-1$, $w=0.5$, the 2 field model matches the single field
predictions to $0.1\%$.  For crossing models  that mimic 
a cosmological constant by having $w_n\approx -1$, e.g. $w_0=-1.15$, $w=0.5$, the deviations
are essentially indistinguishable.}
\label{fig:clmodel}
\end{figure}

\section{Discussion}
\label{sec:discussion}

In Fig.~\ref{fig:time} we show examples of the performance of the model on the 
dark energy observables of
comoving distance $D(z)$, Hubble parameter $H(z)$ and linear growth rate $G(z)$ in the smooth dark
energy
regime (see Eqn.~{\ref{eqn:smooth})
for strong variations in the equation of state $w_a = \pm 0.5$.   To better show the relevant performance
we choose $w_0$ such that $w_n \approx -1$ and 
plot the observables relative to a nearly degenerate fiducial $\Lambda$ model
of
$\Omega_{\rm DE}  = 1-\Omega_m = 0.74$ and $h=0.735$ which is a good fit to the current CMB
data.    In all cases we adjust the $\Omega_{\rm DE}$ of the target model to match
the comoving angular diameter distance to recombination and hence the CMB peak results. 
Note that for $w_a>0$, despite a mismatch in $w$ at very high redshift, all 
observables as a function of redshift are modeled to the $0.1\%$ level.  The performance
for $w_a<0$ is somewhat worse but never exceeds a few times this level.  Thus the two field
model will remain an adequate parameterization until the statistical and systematic errors in the
dark energy observable measurements reaches the sub $1\%$ level.  Mismatches at this
point merely reflect the unavoidable fact that higher order derivatives in $w(a)$ produce observable
effects and our target constant $w_a$ model is itself inadequate.  Even in this
regime the 2 field model can be useful since the parameter $\epsilon_n$ can be 
used to marginalize or probe the second derivative.

The two field model allows one to also calculate the CMB anisotropy.
For a case with no crossing of the divide, e.g. $w_a=0.5$, $w_0=-1$,
the single field model with the exact $w(a)$ matches the 2 field model with the approximate
$w(a)$ to $\sim 0.1\%$, in particular at the low multipoles of the ISW effect (see
 Fig.~\ref{fig:clmodel}).  In addition to the adequate matching of $w(a)$, this 
 indistinguishability is a consequence of setting all of the sound speeds to  $c_e=1$.
 Aside from $w(a)$ the remaining degrees of freedom in
  a dark energy model involve the transition scale between the smooth and clustered
  regimes. 
We also show the predictions for a model that does cross the phantom divide 
$w_a=0.5$, $w_0= -1.15$.  As this model has an equation of state at the normalization 
point of $w_n = -1.025$, it predicts even smaller deviations from the fiducial 
$w=-1$ model with no instabilities in the evolution of the gravitational potential.

The model constructed here permits a self-consistent 
likelihood analysis of dark energy observables
involving both models that cross the phantom divide and
those that differ strongly from $\Lambda$ in the pivot $w_n$.
Although predictions for the former class models differ little
from the currently favored smooth cosmological constant case, the confidence level
at which the latter can be excluded can be affected by a consistent model of
dark energy clustering.  A detailed study of the effect on current cosmological constraints is
beyond the scope of this Brief Report.

If future observations require an evolutionary crossing of the phantom divide, it will
be a good indication that the dark energy contains hidden internal degrees of freedom in
its physical structure.
    
\smallskip
\noindent{\it Acknowledgments:}  We thank C. Armendariz-Picon, S. Carroll, A. Lewis and 
D. Pogosian for useful discussions.
WH was supported by the DOE and the Packard Foundation. This work was begun in the
Aspen Center for Physics and completed at the KICP under NSF PHY-0114422.

\noindent{\it Note added in proof:} After completion of this work, we became
aware of related work by Guo et al [astro-ph/0410654]

\end{document}